\documentclass[
reprint,
aip, 
apl,
numerical, 
11,
twocolumn,showpacs,floatfix,preprintnumbers,amsmath,amssymb]{revtex4-1}
\usepackage{amssymb}
\usepackage{graphicx} % Include FIG.files
\usepackage{psfrag} % Textersetzung in figures
\usepackage{color} 
\usepackage{graphicx}
\usepackage{pstricks-add} % f\"uer das mit geogebra erstellte bild

\newcommand{\lef}{\left(} %linke Klammer
\newcommand{\rig}{\right)} %rechte Klammer

\newcommand{\tdiff}[2]{\ensuremath{    \frac{\mathrm {d}#1}{\mathrm{d} #2}  }}
\newcommand{\refeq}[1]{Eq.\ (\ref{#1})}

\newcommand{\fig}[1]{Fig. (\ref{#1})}

\newcommand{\dd}[0]{\mathrm{d}}

\newcommand{\Roff}[0]{R_\mathrm{off}}
\newcommand{\Ron}[0]{R_\mathrm{on}}

\begin{document}

\title{A Nonlinear HP-Type Complementary Resistive Switch}

\author{Paul K. Radtke and Lutz Schimansky-Geier}  
	
\affiliation{Department of Physics, Humboldt-Universit\"at zu Berlin, Newtonstra{\ss}e 15, 12489 Berlin, Germany}
%\email{radtke@physik.hu-berlin.de}  
%\pacs{05.10.-a,  73.63.-b, 73.50.Fq, 72.80.Tm}%87.10.Mn? PACS, the Physics and Astronomy

\begin{abstract}
Resistive Switching (RS) is the change in resistance of a dielectric under the influence of an external current or electric field. This change is non-volatile, and the basis of  both the memristor and resistive random access memory. In the latter, high integration densities favor the anti-serial combination of two $RS$-elements to a single cell, termed the complementary resistive switch (CRS).  Motivated by the irregular shape of the filament protruding into the device, we suggest a nonlinearity in the resistance-interpolation function, and thereby expand the original HP-memristor. We numerically simulate and analytically solve  this model. Further, the nonlinearity allows for its application to the CRS. 
\end{abstract}

\maketitle

\begin{figure}
\begin{center}
\includegraphics[scale=1.05]{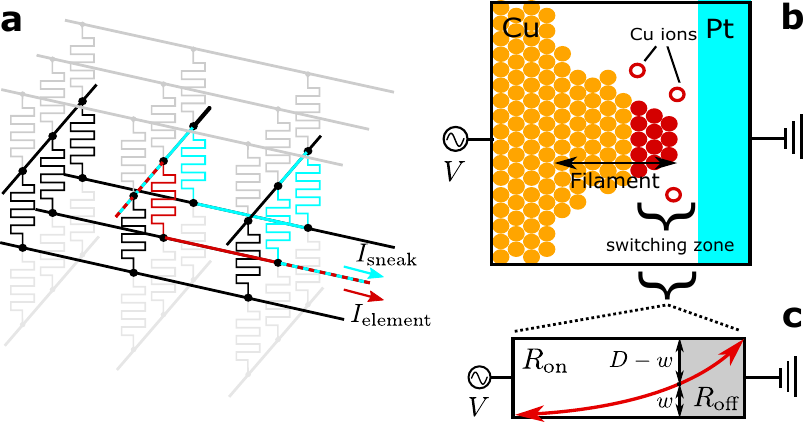}
%\vspace{-.4cm}
\caption{ (\textbf{a}) 3d crossbar nanogrid; current and sneak current in between two wires are shown (cf. \cite{Linn2010}) (\textbf{b}) Schematic of atomic structure in cuprate RS-device. After electroforming a filament of Cu-Atoms protrudes to the other electrode. At its end (red) the switching occurs by Cu-ionizing and disrupting or rebuilding the bridge (\textbf{c}) Nonlinear HP-type memristor, the red line indicates how the position of $w$ determines the relative sizes of the areas with low (white, $\propto (w/D)^p$) and high (grey, $\propto 1-(w/D)^p$) resistance. They correspond to the areas with condensed and free ions of the switching zone in (\textbf{b}) respectively.}
\label{intro}
\end{center}
\end{figure}

\textbf{Introduction.} % Following the pace of myrphys 
In the relentless endeavor of the semiconductor industry to increase performance, the quest for novel technologies and materials is ubiquitous. Among the most promising candidates is resistive random access memory (ReRam). It is expected to provide highly scalable, fast, non-volatile and low cost memory \cite{Waser2009, Jeong2012, Yang2013}. A single such cell consists of a transition metal oxide (TMO) like titanium dioxide \cite{Lee2011, Strukov2008MissingMemristor, Ghenzi2013, Yang2009} or manganites \cite{Chen2005, Ignatiev2006} %or  cuprates \cite{Fujiwara2008} or chalcogenides \cite{Yang2013} 
or an solid electrolyte (electrochemical metalization cells) \cite{Waser2009, Linn2010, Valov2011, Zhuge2015} sandwiched in between two electrodes connected to an external current or voltage. Depending on the amount of charge or applied voltage flux, the elements' resistance toggles between a high and low resistive state, $R_\text{off}$ and $R_\text{on}$. 
%In the crossover between these regions, the resistive switch effect occurs, changing the elements resistance in between a high $R_\text{high}$ and low resistive state $R_\text{low}$. 
Typical implementations for industrial use aim for high density and stack those elements into a 3d nanocrossbar, layered grids of wires with RS cells in between \cite{Vontobel2009, Yang2013, Linn2010}, cf. Fig. (\ref{intro}.\textbf{a}).

Eminent among the problems associated with a high level of integration however is the dissipation of heat. It has been addressed in \cite{Linn2010} by suggesting a  CRS. Therein, two RS elements are combined anti-serially to one memory cell.  Both its logical states have a high resistance, albeit with differing internal states of the constituent elements. One chooses $R^{(1)}_\text{off} \oplus R^{(2)}_\text{on} \widehat{=} 0$ and $R^{(1)}_\text{on} \oplus R^{(2)}_\text{off} \widehat{=} 1$, where the superscript indicates the first and second RS-element respectively. Other combinations have no binary equivalent. Due to the high resistance, the current and associated energy dissipation through the memory cell and occurring sneak paths around it are drastically reduced. As such, the concept has since been picked up by various works, see e.g. \cite{Yu2010, Lee2011, Budhathoki2013, Ambrogio2014, Vourkas2014}.

%%%%%%%%%%%%%%%%%% FIGURE %%%%%%%%%%%%%%%%%%%%%
%%%%%%%%%%%%%%%%%% END %%%%%%%%%%%%%%%%%%%%%%%

In this letter we aim to develop a framework describing such an element based on the original HP-memristor \cite{Strukov2008MissingMemristor}, a device with an doped and an undoped region with different characteristic resistivities. Due to the action of the external driving, dopants begin to drift, shifting the border between both regions, hence altering the resistance. More specifically, the border moves proportionally to the current that flows through the device. Effectively, it can be regarded as two different resistors in a row, whose weights are governed by the border position $w$. Physically, the memory is represented by the state of the device being characterized by either the history of the current or voltage, i.e. by the charge $q(t)=\int^t I(\tau) \dd \tau$ or voltage flux $\phi(t) =\int^t V(\tau) \dd \tau$. The descriptions via $q$ and $\phi$ are equivalent, since for the (ideal) memristor these quantities have a bijective relationship \cite{Chua1971Memristor}, making it possible to express the one as a function of the other \cite{Georgiou2012QuantitativeHysteresis}.

Aforementioned model is most suited to bipolar field effect devices, meaning that the switching can occur with either polarity of the current and that the electric field plays the main role in the process \cite{Strukov2008MissingMemristor, Waser2009}. %, as opposed to effects due to Joule heating \cite{Yang2013} commonly treated by threshold models. 
Prior to normal operation, a conducting filament that permeates the cell is created during an electroforming process \cite{Jeong2012}, it may be build either by electrode ions or oxygen vacancies. The final bridge to the opposite electrode is the active zone where the switching occurs, it is opened and closed depending on the sign of the applied voltage.

Classically, a sheet resistance of a flat conductor with length $D$, breadth $B$ and resistivity $\Ron$ is given by $R = \Ron \cdot D/B$.  We cannot expect the bridge to be a rectangle however, rather it might be some ill-defined wegde-like or pyramidal structure \cite{Waser2009, Ghenzi2013, Yang2013, Zhuge2015}, see Fig. (\ref{intro}.\textbf{b}) or consist of several fused channels \cite{Fujiwara2008}.

Here we suggest an analytically solvable model to reproduce and elucidate the characteristics of the aforementioned device. 
To this end we introduce $(i)$ a nonlinearity in the weighting of the high and low resistance parts.   
$(ii)$ we serially connect two of these novel memristors with counter oriented polarities.  
\newline

%\section{The nonlinear HP-type complementary resistive switch (CRS)}
%\label{strukov-nonlinear}

\textbf{A nonlinear HP-memristor.}
Accounting for the geometric structure of the bridge, we regard the sheet resistance as an integral over the various slices, whose breadth depends on their position. With the boundary $w$ running from $0$ up to the device length $D$, the bridge resistance is given by $R(w)=R_\text{on}\int^w 1/B(x)\dd x$, some general non-linear function. We accordingly set for our model
%In the first modification, we allow the influence of the resistance to be nonlinear, leading to
\begin{align}
\begin{aligned}
R\big(w(t)\big)  =  \Ron \lef \frac{w(t)}{D} \rig^p + \Roff \lef 1 - \lef\frac{w(t)}{D} \rig^p\rig,
\label{strukov-nonlin}
\end{aligned}
\end{align}
with the parameter of nonlinearity $p$. %and the border $w$ moving in between 0 and the device length $D$. 
The corresponding schematics is shown in Fig. (\ref{intro}.\textbf{c}). As for the HP-memristor \cite{Strukov2008MissingMemristor}, the internal variable reacts to the applied current  
\begin{align}
\frac{\mathrm{d}w(t)}{\mathrm{d}t} = & \frac{D}{q_0} I(t).
\label{strukov-w}
\end{align}
Here $q_0$ denotes the total charge accumulated while moving $w=0$ to $w=D$, it is connected to the material constants by $1/q_0 = \mu_D\Ron/D^2$ with the ion mobility $\mu_D$. We remark that this expression needs to be thought of as being supplemented by an additional window function, $f(w)$, which incorporates the limitation of $w$. Once $w$ reaches either $0$ or $D$, it is no further incremented until the motion reverses, neither are the memristor variables $q$ and $\phi$. In the original implementation, $f(w) = \Theta(w) - \Theta(w-D)$ is chosen, where $\Theta(x)$ denotes the Heaviside step function. Several different window functions have been suggested as alterations to the HP-memristor,  differentiating the boundaries movement speed in the bulk and near the electrodes \cite{Joglekar2009, Budhathoki2013, Linn2014,  Biolek2009, Benderli2009}. 

%As will be seen later, such a limitation, or even the logical implementation of such a limitation, is missing in the Rozenberg model, where it leads to a hysteresis in the $q-\phi$-curves. 

The advantage of our nonlinearity modification is that it is simple enough to be solvable. Analogous to the solution of the HP-memristor, we obtain for a single nonlinear element

\begin{align}
%R\big(q(t)\big) = \Ron\lef \frac{w_0}{D} + \frac{q(t)}{q_0} \rig^p + \Roff\lef 1- \lef \frac{w_0}{D} + \frac{q(t)}{q_0}\rig^p\rig,
R\big(q(t)\big) = \Ron\lef  \frac{q(t)}{q_0} \rig^p + \Roff\lef 1- \lef \frac{q(t)}{q_0}\rig^p\rig,
\label{Rofq}
\end{align}
with the initial position $w_0=0$ as boundary condition. Other $w_0$ can be incorporated into \refeq{Rofq} by substituting the charge $q \rightarrow q+q_0w_0/D$.
%with the integration constant $w_0$ denoting the initial boundary position. 

Using Ohm's law $V = R \dd q/ \dd t$, the time derivation operator is put in front of the entire right hand term according to $(a+q)^p \dd q /\dd t = \dd (a+q)^{p+1}/ (p+1) \dd t$, yielding 
\begin{align}
%V\lef q(t), \dot{q}(t) \rig= \tdiff{}{t}\lef \Roff q(t) + \frac{\Ron-\Roff}{p+1} q_0 \lef \frac{w_0}{D} + \frac{q(t)}{q_0} \rig^{p+1} \rig,
V\lef q(t), \dot{q}(t) \rig= \tdiff{}{t}\lef \Roff q(t) + \frac{\Ron-\Roff}{p+1} q_0 \lef \frac{q(t)}{q_0} \rig^{p+1} \rig,
\end{align}
which by time integration gives the voltage flux of the device as a function of the time dependent charge.  With the boundary condition $\phi\lef q=0\rig = 0$ we gain
\begin{align}
\phi\big(q(t)\big) = \Roff q(t) - \frac{\Roff-\Ron}{p+1}q_0\lef \frac{q(t)}{q_0} \rig^{p+1},
\label{qphi-struk-nlin} 
\end{align}
the bijective relationship constitutive for memristive systems.

%%%%%%%%%%%%%%%%%% FIGURE %%%%%%%%%%%%%%%%%%%%%
\begin{figure}
\begin{center}
\includegraphics[scale=0.7]{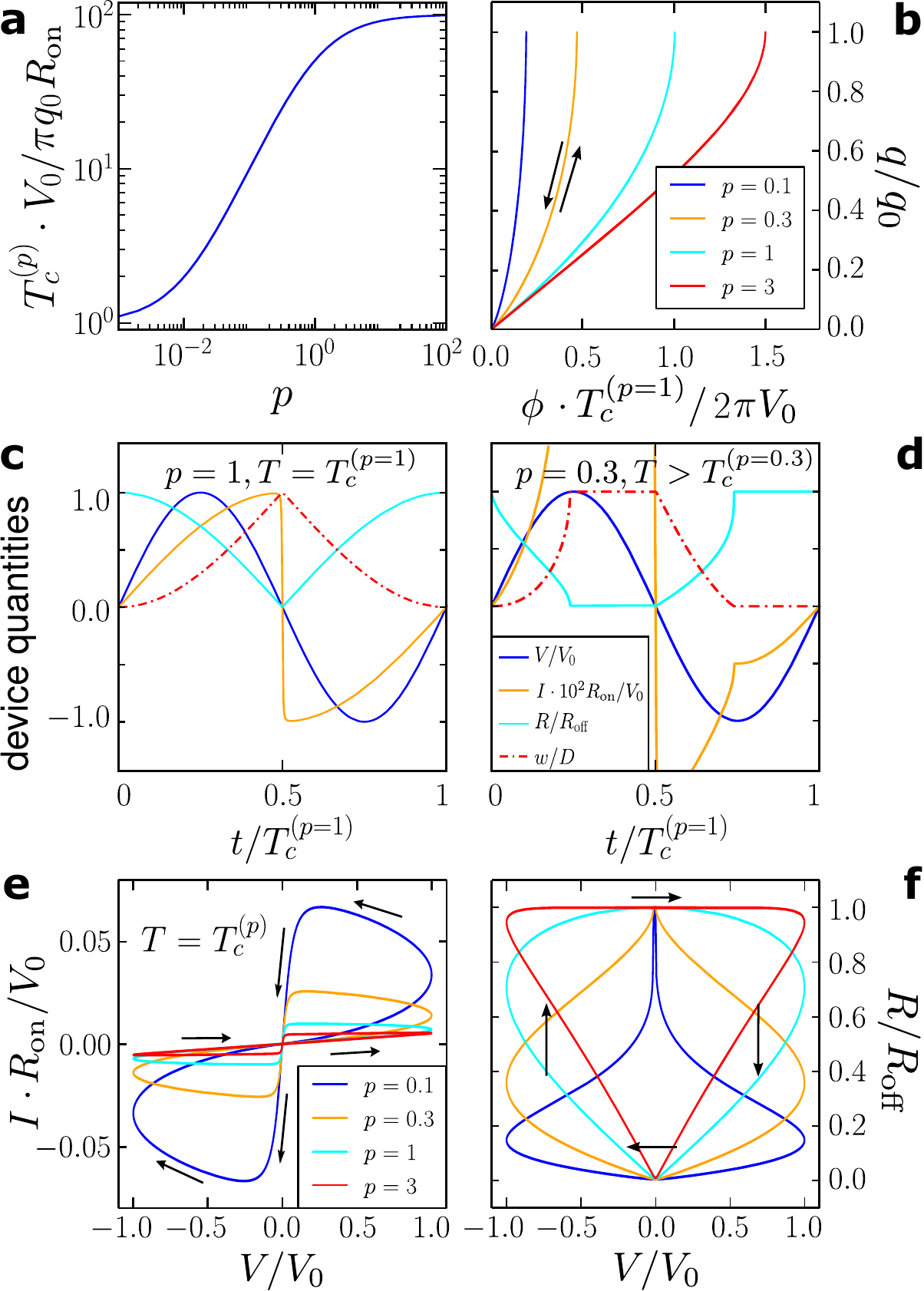}
\caption{Dynamic behavior of a single nonlinear HP-type memristor. (\textbf{a}) Characteristic driving time as a function of the nonlinearity. (\textbf{b}) Associated charge over voltage flux plots. (\textbf{c},\textbf{d}) Time course of the device quantities $R$, $I$, $w$ and $V$  (\textbf{e}) Current over voltage plots, driven by signal with period $T = T_c^{(p)}$ (\textbf{f}) Resistance over voltage plot for various $p$, each for driving period $T = T_c^{(p)}$. Shared legend for (\textbf{e}, \textbf{f}).} % All curves align perfectly with the analytical expressions. 
\label{hp-nonlin}
\end{center}
\end{figure} 
%%%%%%%%%%%%%%%%%% END %%%%%%%%%%%%%%%%%%%%%%%

Let us investigate the behavior of the system. %Reasonable values for the parameters involved are taken from an experiment with titanium dioxide \cite{Strukov2008MissingMemristor}  - 
The parameters are taken from an experiment with titanium dioxide \cite{Strukov2008MissingMemristor}: an ion mobility of $\mu_D = 10^{-14} m^2/V s$, a resistance ratio $\Ron/\Roff = 10^{-2} - 10^{-3}$, so for example $\Ron= 100\Omega$ and $\Roff=20 k\Omega$, and a device size of $D = 10^{-8} m$. This gives a maximal charge of $q_0 = 10^{-4} C$ that the device can hold. We choose a sinusoidal signal for the external driving $V(t) = V_0 \sin\lef 2\pi t/T \rig$ with $V_0 = 1 V$ and $w_0\equiv w(t=0)= 0$. These values are used for all simulations.

\fig{hp-nonlin} depicts the system properties for several nonlinearity parameters $p$ and driving periods $T$. The latter is chosen to be the characteristic period $T_c^{(p)}$ for some $p$, which we define as the value where the device is just completely charged (i.e., from $0$ to $q_0$) during the positive voltage cycle. It follows by $\phi_\mathrm{max} = \int_0^{T_c^{(p)}/2}V(t)\dd t= V_0 T_c^{(p)} / \pi$ and demanding for \refeq{qphi-struk-nlin} $q(T_c^{(p)}/2) = q_0$. With the initial boundary position $w_0=0$ one obtains from \refeq{qphi-struk-nlin}
\begin{equation}
T_c^{(p)} = \frac{\pi q_0}{V_0}\lef \frac{p}{p+1}\Roff + \frac{1}{p+1}\Ron \rig.
\label{characteristic_period}
\end{equation}
Apparently, the characteristic period (cf. Fig. (\ref{hp-nonlin}.\textbf{a})) increases monotonously with $p$, hence for nonlinearities $p \rightarrow 0$ the device switches fastest. This corresponds to the physical case of a filament with a large base getting thinner near the end, presumably also the relevant physical case \cite{Fujiwara2008, Zhuge2015}.

As we see in Fig. (\ref{hp-nonlin}.\textbf{b}) for several nonlinearities, while all describe slightly different memristors, the $\phi-q$ relation remains bijective, as also implied by \refeq{qphi-struk-nlin}. For small $p$ %the resistance drops faster and less time is needed to move the charges in the device. Hence, 
less voltage flux is needed to reach the full charge, which is independent of $p$. \newline

In Fig. (\ref{hp-nonlin}.\textbf{c}) the time courses of the electrical properties are shown for an characteristic driving $T=T_c^{(p=1)}$. At the end of the positive voltage half-cycle $w$ just reaches its extremal position, associated with a minimum of the resistance. In case of the driving times beyond the characteristic period $T=T_c^{(p=1)}>T_c^{(p=0.3)}$, Fig. (\ref{hp-nonlin}.\textbf{d}), $w$ saturates before that, and consequently the current spikes as the voltage still increases.

%In the $V-I$ diagram for the $T_c^{(p=1)}$ Fig. (\ref{strukov-nonlin}.\textbf{c}), we see that for $p \geq 1$ the curves retain their symmetry. Since the smaller $p$ is, the device has a smaller resistance for a longer time and correspondingly the maximal current flowing through the system until saturation for $q_\mathrm{max}^{(p=1)}$ does not suffice to lead the system with $p>1$ to saturation, it only has a small hysteresis loop. Systems with $p<1$ on the other hand have a lower resistance, and therefore the same voltage flux leads to a larger current, the reply curve loses its symmetry. For positive voltages, the resistance reaches its minimum before the voltage peaks and hence and $I$ significantly increases. This behavior is also visible in the time courses of the electric quantities Fig. (\ref{strukov-nonlin}.\textbf{c}), for $p=0.3$ the device is driven with a period above its characteristic value, $T=T_c^{(p=1)}>T_c^{(p=0.3)}$ and hence the current explodes while $w$ no longer changes. 

In the $V-I$ diagram for $T=T_c^{(p)}$ (Fig. (\ref{hp-nonlin}.\textbf{e})), i.e. at each respective characteristic time, we see the pinched hysteresis loops that are the fingerprint of memristive systems \cite{Chua2011}. The area enclosed by the hysteresis cycles  increases with smaller $p$. In the $R-V$ diagram Fig. (\ref{hp-nonlin}.\textbf{f}) this shows by a curve that transforms from being triangularly shaped to one resembling an onion, with the average resistance falling with smaller $p$. 
%In the $R-V$ diagram for at each devices characteristic driving $T=T_c^{(p)}$ Fig. (\ref{strukov-nonlin}.\textbf{d}) we see that the curve transforms from being triangularly shaped to one resembling an onion, with the average resistance again falling with the smaller $p$ is. In the $V-I$ diagram this would correspond to decreasing areas enclosed by the hysteresis loops.

\begin{figure}
\begin{center}
\includegraphics[scale=1.4]{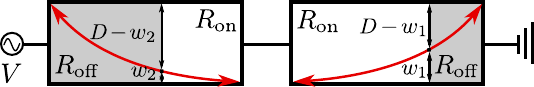}
\caption[Setup to realize bipolar resistive switching with two nonlinear Strukov-type Memristors]{
 Setup to realize bipolar resistive switching with two nonlinear HP-type memristors in anti-serial order. }
\label{strukov-bipolar}
\end{center}
\end{figure}

\textbf{The CRS composed of two nonlinear HP-memristors. }
In the next iteration, we consider two memristors connected in a row with counter oriented polarities. This setup is rendered possible only by the nonlinearity in the $w$-dependence of the resistance (cf. \refeq{strukov-nonlin}), since otherwise (i.e. $p=1$) both elements' behavior just cancels each other out. Conceptually, this represents realizations with both an implementation of two RS-elements with one active zone each and one element that has active zones near both electrodes \cite{Yang2009, Linn2010, Lee2011}. Let the two domain boundaries be denoted by $w_1$ and $w_2$. The resistance is now simply given by the sum of the individual contributions,
\begin{align}
\begin{aligned}
%V(t)  = \bigg[& 
%R_{\mathrm{on}} \lef \lef \frac{w_1(t)}{D} \rig^p + \lef \frac{w_2(t)}{D}\rig^p \rig\\+ 
%& R_\mathrm{high} \lef \lef 1-\frac{w_1(t)}{D}\rig^p + \lef 1 - \frac{w_2(t)}{D}\rig\rig^p \rig \bigg] I(t),
R_\mathrm{CRS}\big(w_1(t), w_2(t)\big) = R\big(w_1(t)\big) + R\big(w_2(t)\big).
\label{resistance-crs}
\end{aligned}
\end{align}
The response of each memristor to the current flow is not altered, i.e. still governed by \refeq{strukov-w}, albeit with a minus sign for the second element. This layout is illustrated in Fig. \ref{strukov-bipolar}. Since we have two elements in a row, the voltage drop at each is proportional to its resistance and in extension, so is the voltage flux $\phi$. 

%%%%%%%%%%%%%%%%%% FIGURE %%%%%%%%%%%%%%%%%%%%%
%%%%%%%%%%%%%%%%%% END %%%%%%%%%%%%%%%%%%%%%%%

\begin{figure*}
\begin{center}
\includegraphics[scale=0.65]{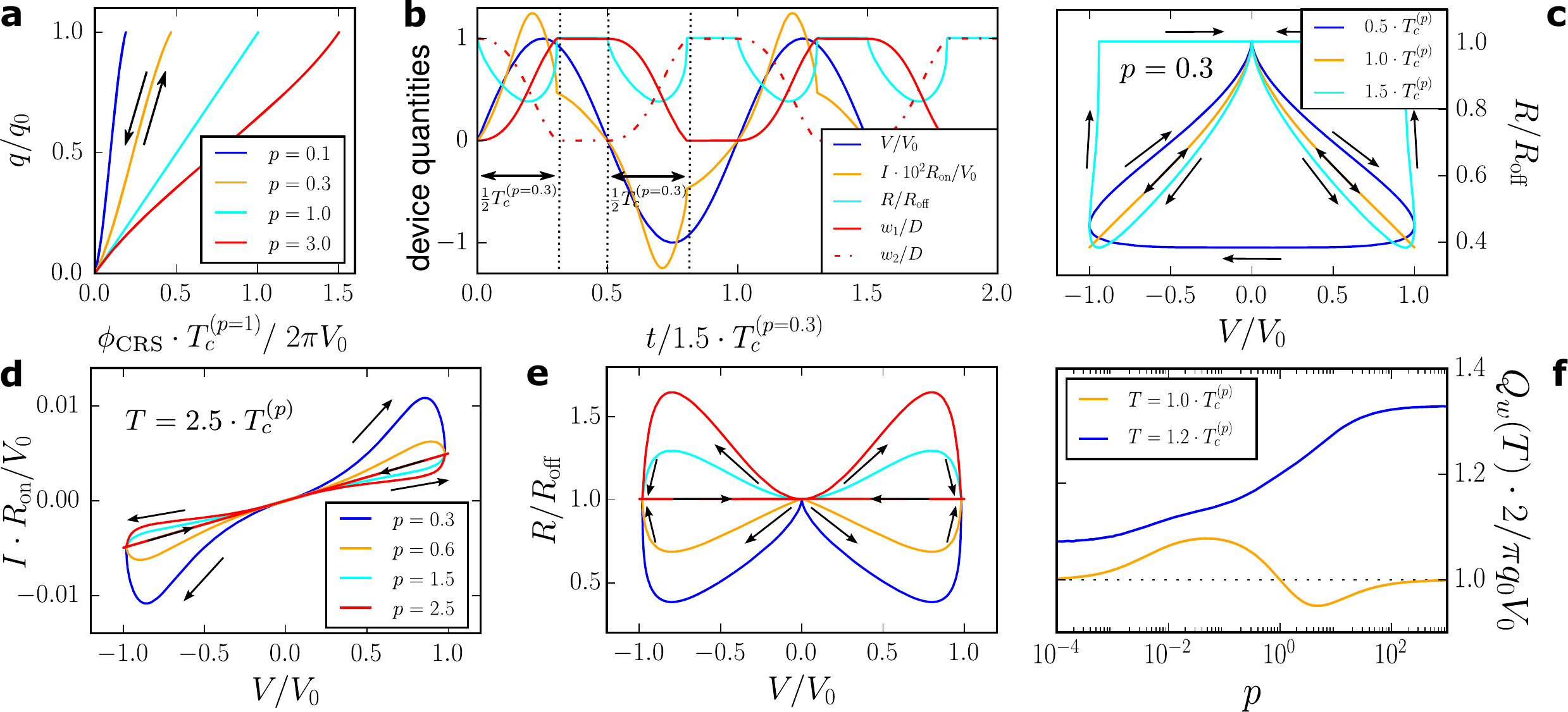}
\caption{Dynamic behavior of CRS composed of nonlinear HP-type memristors. (\textbf{a}) Charge over voltage flux for several nonlinearity parameters.  (\textbf{b}) Time dependence of system quantities for $p=0.3$ and driving period $1.5\cdot T_c^{(p=0.3)}$, the two half cycles of the corresponding characteristic period are indicated by the double headed arrow. (\textbf{c}) Current over voltage plot for $p=0.3$ and various driving periods. % that are various multitudes of $T_c^{(p=0.3)}$.
%the characteristic time 
(\textbf{d}) Current over voltage and (\textbf{e}) resistance over voltage plots for various nonlinearities $p$, each driven by signal with period $2.5 \cdot T_c^{(p)}$, shared legend for (\textbf{d}, \textbf{e}). %a the same multitude of their respective characteristic period, $2.5 \cdot T_c^{(p)}$. 
(\textbf{f}) Heat dissipation $Q_w$ during one switching operation.}
\label{strukov-2memristors}
\end{center}
\end{figure*}

For the first element $\phi_1(q)$ is given by \refeq{qphi-struk-nlin}, whereas for the second we have to incorporate the reversed polarity as well as the boundary conditions $w^{(2)}_0=D$ and $\phi(q=0)=0$, yielding the flux $\phi_2(q) = \phi(q_0) - \phi(q_0-q)$.  Addition of both gives the total flux,
\begin{align}
\label{qphi-bi}
 \phi_\text{CRS}\big(q(t)\big) = & 2\Roff q(t) - \frac{\Roff-\Ron}{p+1}q_0 \times \\
& \lef \lef \frac{q(t)}{q_0}\rig^{p+1}+ 1 - \lef 1 - \frac{q(t)}{q_0} \rig^{p+1} \rig. \nonumber
\end{align} 
The term in bracelets of  \refeq{qphi-bi} has a functional dependence on the charge  $q$ of the order $p$. Hence for $p=1$, this gives $\phi_\text{ CRS }(t) = (\Roff + \Ron)q(t)$, the linear $q-\phi$ relation of a constant resistor. We note that from the the functional form of \refeq{qphi-bi} that the $\phi-q$ relation remains bijective. This is straightforward, since the combination of memristors must yield another one \cite{Chua1971Memristor}. As expected, each nonlinearity parameter describes a different memristive system with a relation complying to \refeq{qphi-bi}, Fig. (\ref{strukov-2memristors}.\textbf{a}). For smaller $p$ a smaller voltage flux $\phi$ is needed to charge to memristor. We can hence see it as more reactive with smaller $p$, or put differently, its average resistance over a voltage cycle is lower, cf. Fig. (\ref{strukov-2memristors}.\textbf{e}).

In comparison with the characteristic period for one nonlinear element \refeq{characteristic_period}, we find for the CRS it simply doubles, since the voltage flux counts separately for each of the elements, hence 
\begin{equation}
T_c^{(p)} = \frac{2 \pi q_0}{V_0}\lef \frac{p}{p+1}\Roff + \frac{1}{p+1}\Ron \rig.
\label{cms-tc}
\end{equation}

%%%%%%%%%%%%%%%%%% FIGURE %%%%%%%%%%%%%%%%%%%%%
%%%%%%%%%%%%%%%%%% END %%%%%%%%%%%%%%%%%%%%%%%

In Fig. (\ref{strukov-2memristors}.\textbf{b}) some system reply properties for this dynamics are shown for a single value of $p$. Within each half cycle of the driving, somewhere before the voltage peak the resistance drops once, accompanied by a spike in the current. This is due to the inherent symmetry of the device. Assuming that the first element is in the off-state and the second in the on-state, a positive voltage will now move the first in the on-state and the second in the off-state, with both configurations having the same resistance.  However, in the course of that both will be in an intermediate state in which the resistance drops (for $p < 1$). The same happens the other way around with the negative voltage half cycle. Hence the doubling of the frequency of the reply quantity $R$. It is mirrored in the $R-V$-plots (\ref{strukov-2memristors}.\textbf{c},\textbf{e}) by the axis symmetry with respect to the driving voltage.

Further investigating the resistance course for various driving periods (\ref{strukov-2memristors}.\textbf{c}) reveals a behavior resembling a single element for $T < T_c^{(p)}$, cf. Fig (\ref{strukov-nonlin},\textbf{e}). For the characteristic driving $T=T_c^{(p)}$ on the other hand the course is given by two bidirectional paths without hysteretic spread. Finally, for $T>T_c^{(p)}$ a two-legged figure appears, mirroring the frequency doubling in the reply properties already noticed in the $R(t)$-plot. It has clearly distinguished zones of high and low resistance. In compliance with experimental results for the bipolar resistive switching \cite{Chen2005, Ignatiev2006, Rozenberg2010a}, this is the case of interest.

The impact of several nonlinearities on the $V-I$- and $V-R$-plots is shown in Fig. (\ref{strukov-2memristors}.\textbf{d}-\textbf{e}). We observe the pinched hysteresis loops for all. In comparison to a single element Fig. (\ref{hp-nonlin},\textbf{e}), not only are the curves symmetric to the point of origin, but also their directionalities. The area enclosed by the hysteresis loops drops for increasing $p$.  In the resistance plots this shows by an inversion of the legs, the symbolic legs point skywards.

Finally, we turn our attention to the dissipative loss accumulated by the system. To that end, we regard the Joule heat $Q_w$ generated during a switching operation $0 \leftrightarrow 1$, $Q_w = \int^{T/2} P(t)\dd t = \int^{T/2} V^2(t)/R(t)\dd t$. Driven by the characteristic time, the result is shown in Fig. (\ref{strukov-2memristors}.\textbf{f}). Except for a slight variation in the middle, the heat loss for vanishing and huge $p$ approach the same value. This asymptotic can be understood in view of the fact that for $p\rightarrow 0$, the resistance goes to $2\Ron$ (cf. Eqs. (\ref{strukov-nonlin}), (\ref{resistance-crs})), whereas the characteristic time \refeq{cms-tc} approaches $2\pi q_0 \Ron/V_0$. For $p\rightarrow \infty$ we have $R_\text{cms} \rightarrow 2\Roff$ and $T_c^{(\infty)} = 2\pi q_0 \Roff/V_0$.  Hence for both cases the integral evaluates to $Q_w = \pi q_0 V_0/2$.  Given the size of the difference in $Q_w$, the nonlinearity is fairly secondary to the heat loss. For longer driving periods $T=1.2\cdot T_c^{(p)}$ on the other hand, the heat loss increases continuously with $p$ and small $p$ seem most beneficial. 

\textbf{Conclusion.} In this work we have developed a nonlinear alteration of the original HP-memristor, that remains analytically treatable. The nonlinearity reflects the irregularity of the filament structure that bridges both electrodes. For a single element we are able to reproduce all the characteristic curves and predict an increased efficiency in the switching times with decreasing nonlinearities.

In the second part, we extended this model by the anti-serial combination of two such elements. We are now able to reconstruct the reply curves of the CRS without losing analytical tractability and we can predict the behavior of the CRS with a memristive model. 
Thereby, the cases $p<1$ and $T>T_c^{(p)}$ are of most physical relevance, for they reproduce the two-legged structures seen in experimental data for anti-serial bipolar switching \cite{Chen2005, Rozenberg2010a}. Further, here we have the advantage of faster switching and reduced dissipative heat loss in case of driving times beyond $T_c^{(p)}$.

\textbf{Acknowledgments.} This paper was developed within the scope of the IRTG 1740 funded by the DFG. We thank B. Sonnenschein for comments on the manuscript.

\bibliography{../bibliography/library}

\end{document}